\documentstyle[12pt,epsf]{article}
\def\tf#1#2{{\textstyle{#1 \over #2}}}
\def\beq{\begin{equation}}
\def\eeq{\end{equation}}
\def\bea{\begin{eqnarray}}
\def\eea{\end{eqnarray}}
\def\bef{\begin{figure}}
\def\enf{\end{figure}}
\def\A{{\bf A}}
\def\S{{\bf S}}
\def\C{{\bf C}}
\def\Z{{\bf Z}}
\def\R{{\bf R}}
\def\N{{\bf N}}

\def\i0{{i}}
\def\j0{{j}}
\def\T11{{\bf T}^{1,1}}
\def\AdS5{{\bf AdS}_5}
\def\P1{{\bf P}^1}
\def\tCkl{{\widetilde{{\cal C}_{kl}}}}
\def\hCkl{{\widehat{{\cal C}_{kl}}}}
\def\tX5{{\widetilde{X_5}}}

\def\ba{\begin{array}}
\def\ea{\end{array}}
\def\bce{\begin{center}}
\def\ece{\end{center}}

\newcommand{\drawsquare}[2]{\hbox{%
\rule{#2pt}{#1pt}\hskip-#2pt%  left vertical
\rule{#1pt}{#2pt}\hskip-#1pt%  lower horizontal
\rule[#1pt]{#1pt}{#2pt}}\rule[#1pt]{#2pt}{#2pt}\hskip-#2pt%  upper horizontal
\rule{#2pt}{#1pt}}% right vertical

\newcommand{\fund}{\raisebox{-.5pt}{\drawsquare{6.5}{0.4}}}%  fund
\newcommand{\antifund}{\overline{\fund}}

\begin{document}
\begin{titlepage}
\rightline{BROWN-HET-1215}
\rightline{hep-th/0003183}
\vskip 2cm
\centerline{ \Large\bf{Renormalization Group Flows on D3 branes}}
\vskip 1cm
\centerline{\Large\bf{at an Orbifolded Conifold}}
\vskip 1cm
\centerline{\sc Kyungho Oh$^{a}$ and Radu
Tatar$^{b}$}
\vskip 1cm
\centerline{ $^a$ Dept. of Mathematics, University of
Missouri-St. Louis,
St. Louis, MO 63121, USA }
\centerline{{\tt oh@math.umsl.edu}}
\centerline{$^b$ Dept. of Physics, Brown University,
Providence, RI 02912, USA}
\centerline{\tt tatar@het.brown.edu}
\vskip 2cm
\centerline{\sc Abstract}
We consider D3-branes at an orbifolded
conifold whose horizon ${X_5}$ resolves into
a smooth 
Einstein manifold which joins several copies of ${\bf T}^{1,1}$. 
We describe in details the resolution 
of the singular  horizon ${X_5}$ and describe different types of two-cycles 
appearing in the resolution. 
For a large number of D3 branes, 
the AdS/CFT conjecture becomes a duality between type IIB string theory
on ${\AdS5 } \times {X_5} $ 
and the ${\cal N} = 1$ field theory living on
the D3 branes. We study the fractional branes as small perturbations of the
string background and we reproduce
the logarithmic flow of field theory couplings by studying fluxes of NS-NS and R-R two forms
through different 2-cycles of the resolved horizon.

\def\today{\ifcase\month\or
January\or February\or March\or April\or May\or June\or
July\or August\or September\or October\or November\or December\fi,
\number\year}
\vskip 1cm
\end{titlepage}
%****************************************************
%*****************************************************
%*****************************************************
\newpage
\section{Introduction}
In the last years we have seen increasing evidences that string/M theory on
AdS spaces are 
dual to large $N$ strongly coupled gauge theories \cite{mal,mal1}. 
The most extensively studied  cases  are the dualities
 between Type IIB string theories on
${\bf AdS_{5}} \times M_5$  for
positively curved five dimensional Einstein manifolds and
 large $N$ strongly coupled four dimensional
conformal gauge theories. 
The simplest example is  ${\bf AdS_{5}} \times {\bf S^5}$. In this case, 
the dual field theory is
${\cal N} = 4$ supersymmetric gauge theory. 
In ~\cite{kac,law}, field theories with less supersymmetry have been studied 
as duals 
to string theory on orbifolds of  ${\bf S^5}$.
In \cite{kw}, Klebanov and Witten studied the Einstein manifold
${\bf T}^{1,1} = (SU(2) \times SU(2))/U(1)$.  This was
the first example of five dimensional spaces
 which are not  orbifolds of ${\bf S^5}$. 
Type IIB string theory compactified on this manifold is dual to an
${\cal N} = 1$ superconformal $SU(N) \times SU(N)$ gauge theory with a quartic 
superpotential for bifundamental fields.  In the T-dual picture,
D3 branes probing a metric cone over ${\bf T}^{1,1}$ (which is 
the conifold) is either
a brane configuration
with D4 branes together with orthogonal NS branes \cite{ura,dm} 
or a brane box
with D5 branes together with orthogonal NS branes~\cite{hz,hu}.
Other results on the
conifold or the quotients of the  conifold and their field theory duals 
were obtained in 
\cite{kn,afhs,gk,gub1,mr,lopez,unge,karch,kw1,gk,gns,ot1,ura1,oy}.

Another exciting development was study of the
gravity dual of the field theory Renormalization group flow. The main point is that the
radial coordinate of ${\AdS5}$ has the natural interpretation as an 
energy scale in field theory.  Thus it becomes natural to consider 
type IIB supergravity interpolating solutions where the metric and the fields
depend on the radial coordinate, and to interpret these solutions as 
RG flows in the dual field theory. 
Many ideas have emerged concerning different aspects of
supersymmetric and non-supersymmetric RG flows for four dimensional theories 
\cite{zaf1,dz,fgpw,gub2,bvv,sch,gub,kto,lowe}. 

Because the AdS/CFT involves the full string theory we should be able 
to go beyond the Supergravity approximation.  
In \cite{kn} Klebanov and Nekrasov have studied 
the gravity duals of fractional branes in supersymmetric conifold and orbifold theories where
the presence of the fractional branes breaks the conformal invariance and 
introduces an RG flow in field theories.
One of the models considered in \cite{kn} involves
 a large number of D3-branes at a conifold singularity 
whose near-horizon is  a $\AdS5 \times \T11$ background 
and the fluxes of $B^{NS}$ and $B^{RR}$ forms through
the blow-up 2-cycle determine a difference in the  coupling constants of 
the two group factors appearing on the
world-volume of the D3 branes at singularities. The supergravity
equations of motion together with the specific formulas for the 2-forms and 3-forms on ${\bf T}^{1,1}$ 
give a solution which reproduces the logarithmic flow of field theory beta function.
Previous results were obtained in type 0B string where effective action
uncertainties occur  \cite{kt1,kt2,kns,ar1,ar2}. 
In all the studies of RG flow from AdS/CFT, the RG flow was determined by turning
on different operators in the field theory which break the conformal
invariance. In supergravity this means  turning on some of the supergravity
scalar fields. Another way to break the conformal invariance is to
introduce the twisted sectors of string theories.

In this paper we go one step further and study D3 branes on 
an orbifolded conifold which is the quotient of the conifold
by $\Z_k \times \Z_l$. Now the horizon $X_5$ will be  singular along
two disjoint, but linked circles and we need 
to resolve the singularities in order to obtain a smooth Einstein manifold 
$\tX5$.
 We completely describe the resolution of
the orbifolded conifold itself  in two steps and discover that
there are $kl$  isolated conifold singularities after the first step of the
resolution. 
After the resolution we find  a smooth Einstein manifold $\tX5$ containing
 $kl+k+k-2$ two-cycles. As $\tX5$ approaches the exceptional fiber of the first resolution,
$kl$ cycles are vanishing into the singular points, and $k+l-2$ cycles
 deform to
cycles in the fiber which  separate the two  circles of singularities.
Near each singular point, $\tX5$ can be approximated by ${\bf T}^{1,1}$ and
$kl$ two-cycles of $\tX5$ come from these ${\bf T}^{1,1}$'s.
We then consider a large number of D3 branes probing this singularity, which
corresponds to a brane box with D5 branes and orthogonal NS branes 
via T-duality~\cite{hz,hu,karch}. 
The D5 branes wrapped on 2-cycles of $\tX5$ vanishing into the singular points of the partially
resolved orbifolded conifold are the fractional D3 branes.  
We study the $B^{NS}$ and $B^{RR}$ fluxes through different 2-cycles of $\tX5$
which give rise to  a logarithmic flow for the field theory coupling constants.
This
 agrees with the field theory expectations for the RG flow.

In section 2 we study the geometry of the orbifolded conifold
and  describe how to  
obtain a smooth horizon from the singular horizon 
${\bf T}^{1,1}/{\Z_k \times \Z_l}$. 
We also identify the different fractional D3 branes in the 
singularity picture with different components of
brane interval or brane box configurations obtained by T-dualities. 
In section 3 we describe the supergravity dual to the field theory Renormalization Group flows. 
%%%%%%%%%%%%%%%%%%%%%
%%%%%%%%%%%%%%%%%%%%%%
%%%%%%%%%%%%%%%%%%%%%%
\section{Geometry and Brane Configurations of Orbifolded Conifolds}
\setcounter{equation}{0}
In this section we study the geometry of the orbifolded conifolds ${\cal C}_{kl}$ in detail and we 
make connections with brane configurations obtained by T-dualities. In particular, we study the
resolutions of the orbifolded conifolds ${\cal C}_{kl}$ and the associated  fractional branes in the T-dual picture. 
At the end of the section, we describe
 the homological cycles of the resolved horizon of ${\cal C}_{kl}$, and
 thus extends the results of \cite{co,kw}. This play an important role 
in the study of fluxes in the next section.

Consider a singular Calabi-Yau threefold $Y_6$  which is a metric cone over
a five dimensional Einstein manifold $M_5$. Then the metric near the apex of
the cone will be
\begin{equation}
ds_{Y_6}^2 = d r^2 + r^2 ds_{M_5}^2.
\end{equation}
Here the apex is located at $r=0$ and $M_5$ is called the horizon of
the cone $Y_6$.
If $N$  parallel  D3 branes are placed at the apex of
the cone  $Y_6$, the resulting ten dimensional spacetime has the metric
\begin{equation}
ds^2 = R^2 \left [ {r^2\over R^4} (-dt^2 + dx_1^2 + dx_2^2 + dx_3^2) +
{dr^2\over r^2}+ds_5^2 \right ]\ ,\qquad R^4\sim g_s N (\alpha')^2.
\end{equation}
The near-horizon ($r \to 0$) limit of the geometry is 
${\bf AdS_{5}}\times M_5$. Type IIB theory on this background is conjectured to
be
dual
to the conformal limit of the field theory on the D3 branes.

In \cite{kw},  an example of such duality has been discussed  in the case
of a conifold
whose horizon  is ${\bf T}^{1,1}= (SU(2)\times SU(2))/U(1)$. The conifold is a three dimensional hypersurface singularity in 
$\C^4$ defined by:
\bea
\label{conieqn}
{\cal C}: \quad z_1z_2 -z_3z_4 = 0.
\eea
The conifold can be realized as a holomorphic quotient of $\C^4$
by the $\C^*$ action given by \cite{kw}
\bea
(A_1, A_2,B_1, B_2)\mapsto (\lambda A_1, \lambda A_2,\lambda^{-1} B_1,
\lambda^{-1} B_2)\quad\mbox{ for }\lambda \in \C^*.
\eea
Thus the charge matrix is the transpose of $Q^{'}
=(1,1,-1,-1)$ and $\Delta=\sigma$ will be a convex polyhedral cone
in $\N^{'}_{\R}=\R^3$ 
generated by $v_1, v_2, v_3, v_4 \in \N^{'}=\Z^3$  where
\bea
v_1=(1,0,0), \quad v_2=(0,1,0),\quad  v_3=(0,0,1),\quad
v_4=(1,1,-1).
\eea
The isomorphism between the conifold ${\cal C}$ and the holomorphic
quotient is given by
\bea
\label{act}
z_1=A_1B_1, \quad z_2=A_2B_2, \quad z_3=A_1B_2, \quad z_4=A_2B_1. 
\eea
To identify the horizon  from this point of view, note that 
we can divide by the scaling $z_i \to sz_i$ (with real positive $s$) by setting
$|A_1|^2 + |A_2|^2 = |B_1|^2 + |B_2|^2 =1$. This gives us 
${\bf S}^3 \times {\bf S}^3 = SU(2) \times SU(2)$. Then dividing by
the $U(1)$ action
\bea
(A_1, A_2,B_1, B_2)\mapsto (e^{i\alpha} A_1, e^{i\alpha} A_2,e^{-i\alpha} B_1,
e^{-i\alpha} B_2),
\eea
we obtain ${\bf T}^{1,1} = (SU(2) \times SU(2))/U(1)$.

The five dimensional manifold ${\bf T}^{1,1}$ has 2-cycles and 3-cycles.
Besides the D3 branes orthogonal to ${\bf T}^{1,1}$, there are wrapped
D3 branes over the 3-cycles of ${\bf T}^{1,1}$ (which
correspond to ``dibaryon'' operators \cite{gk}) and wrapped D5 branes
over 2-cycles of ${\bf T}^{1,1}$ (which   to correspond to
domain walls in ${\bf AdS_{5}}$ \cite{gk} and to fractional D3 branes 
\cite{ddg,d,dm1}). Because ${\bf T}^{1,1}$ is ${\bf S^2 \times S^3}$, we can
identify the 2-cycle with ${\bf S^2}$ and the 3-cycle with ${\bf S^3}$. These two cycles are
orthogonal so the D3 brane wrapped on ${\bf S^3}$ is orthogonal to the D5 brane
wrapped on ${\bf S^2}$, therefore when they cross each other a fundamental
string is created as explained in \cite{bdg,dfk,hw} and the gauge group
becomes $SU(N +1) \times SU(N)$. 

\begin{figure}
\centering
\epsfxsize=3.5in   
\hspace*{0in}\vspace*{.2in}
\epsffile{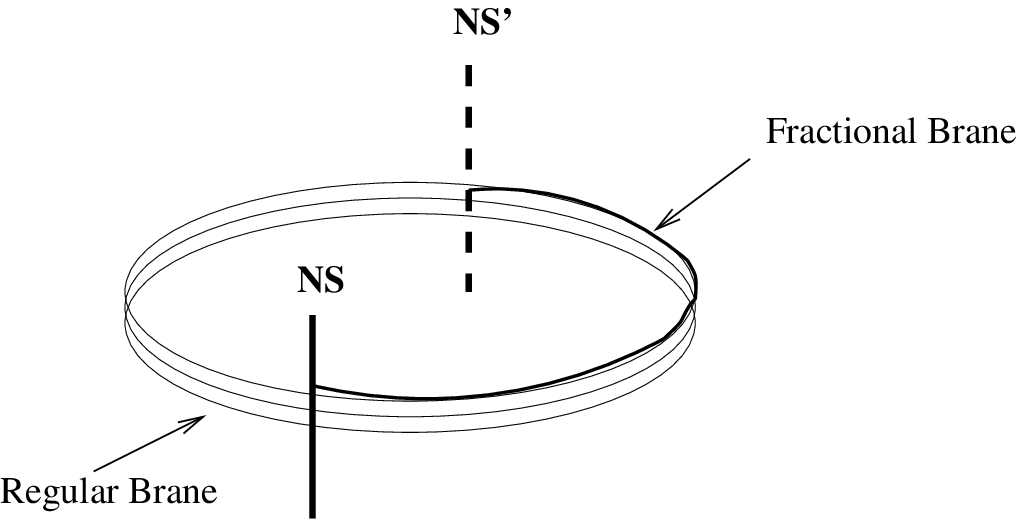}
\caption{An Interval Model} 
\label{frac} 
\end{figure}

The geometrical picture is T-dual to different types of brane configuration. By one T-duality one can obtain the brane interval 
picture with D4 branes wrapped on a circle and by two T-dualities one obtains the brane box picture with D5 branes wrapped on
a 2-torus. The fractional branes have also been identified in  the brane interval picture in ~\cite{dm1}. The idea was to 
interpret the conifold (\ref{conieqn}) as a ${\bf C}^*$ fibration over the
${\bf C}^2$ parameterized by $z_3, z_4$.  By performing T-duality along the $U(1)$-orbit
in the ${\bf C}^*$-fiber, we obtain from the degenerate fibers $z_1 =0$ and $z_2 =0$, 
two NS fivebranes extended, say, $x^0x^1x^2x^3x^4x^5$ and $x^0x^1x^2x^3x^8x^9$
 directions which we denote by NS and NS' branes. 
The D3 branes located at the singular point transform into D4 branes wrapping a circle
which is transverse to the NS fivebranes. $\T11$ has a $U(1)$-fibration over
${\P1}\times {\P1}$ and a two cycle ${\bf S^2}$ of
${\T11}$ can be  identified to the difference of two homologically distinct spheres coming from ${\P1}\times {\P1}$. 
After identifying ${\P1}\times {\P1}$ with the exceptional locus in the full resolution of 
the conifold,  D5 brane wrapping the two cycle ${\bf S^2}$ will transform as a D4 brane
wrapping on one interval between two NS-branes. This is a fractional brane
in the interval model (Figure \ref{frac}).

%%%%%%%%%%%%%%%%%%%%%%%%
%%%%%%%%%%%%%%%%%%%%%%%%%

One of the goals of this paper is to study
 the fractional branes in the brane box model for a quotient of the conifold.
To do this, we start by  taking a further quotient of the conifold ${\cal C}$ by a discrete group
$\Z_k \times \Z_l$. Here $\Z_k$ acts on $A_i, B_j$ by
\bea
\label{zk}
(A_1, A_2, B_1, B_2) \mapsto 
(e^{-2\pi i/k} A_1, A_2, e^{2\pi i/k}B_1, B_2),
\eea
and $\Z_l$ acts by
\bea
\label{zl}
(A_1, A_2, B_1, B_2) \mapsto 
(e^{-2\pi i/l} A_1, A_2, B_1, e^{2\pi i/l}B_2).
\eea
Thus they will act on the conifold ${\cal C}$ by
\bea
\label{xy}
(z_1,z_2,z_3,z_4) \mapsto (z_1, z_2, e^{-2\pi i/k}z_3, e^{2\pi i/k}z_4)
\eea
and
\bea 
\label{uv}
(z_1,z_2,z_3,z_4) \mapsto (e^{-2\pi i/l}z_1, e^{2\pi i/l}z_2, z_3, z_4).
\eea
Its quotient is  called  the orbifolded conifold (or 
the hyper-quotient of the conifold) and denoted by ${\cal C}_{kl}$.

Note that the action (\ref{zk})
 leaves a complex two space $A_1 = B_1 =0$ fixed in $\C^4$
and this is isomorphic to $\C^1$ given by $z_1 = z_3 =z_4 =0$
on the conifold ${\cal C}$ after dividing by the $U(1)$ action.
Similarly, the action (\ref{zl}) leaves fixed a complex two space $A_1 = B_2 =0$ in $\C^4$
and it is isomorphic to $\C^1$ given by $z_1 = z_2 =z_3 =0$ on the conifold ${\cal C}$ after dividing by the $U(1)$ action.
Furthermore, the action (\ref{zk}) descends to the horizon $\T11$ and leaves the following circle fixed:
\bea
|z_2|^2 =1,\quad z_1 = z_3=z_4=0
\label{circle1}
\eea
or equivalently $|A_2|^2 = |B_2|^2 =1, A_1 = B_1 =0 \pmod {U(1)}$.
Similarly, the action (\ref{zl}) leaves the following circle fixed
\bea
|z_4|^2 =1,\quad  z_1=z_2= z_3 =0
\label{circle2}
\eea
or equivalently  $|A_2|^2 = |B_1|^2 =1, A_1 = B_2 =0  \pmod {U(1)}$.
Hence the horizon $X_5 := \T11/\Z_k \times \Z_l$ of the orbifolded conifold
is singular 
along these two circles. These two circles are separated but linked. The
horizon $X_5$ has $\A_{k-1}$ singularity along the circle (\ref{circle1})
and $\A_{l-1}$ singularity along the circle (\ref{circle2}).
String theory in the back ground ${\AdS5} \times X_5$ has massless fields which are localized along these two linked  circles. 
As discussed in \cite{hu}, these massless fields are the twisted modes and they propagate on 
${\AdS5 \times \S^1} \sqcup {\AdS5 \times \S^1} $ where
${\S^1}\sqcup\S^1$ are the  circles of singularities. As we shall see below, there are $k+l-2$ 2-cycles which separate
these two circles. The fluxes of the NSNS and RR two forms through these cycles give rise to scalars which live
in the ${\AdS5 \times \S^1}$ space and are the same as the scalars introduced in section 3 of \cite{kn}.

To put the actions (\ref{act}), (\ref{zk}) and (\ref{zl}) on an equal footing,
consider the over-lattice $\N = \N^{'} + \frac{1}{k}(v_3-v_1) + \frac{1}{l}(v_4 -v_1)$.
Now the lattice  points $\sigma \cap \N$ of $\sigma$ in $\N$
are generated by $(k+1)(l+1)$ lattice points as a semigroup. The discrete group $\Z_k \times \Z_l \cong 
\N / \N^{'}$ will act on the conifold $\C^4 // U(1)$ and its quotient
will be the symplectic reduction $\C^{(k+1)(l+1)} // U(1)^{(k+1)(l+1)-3}$. 
The new toric 
diagram for ${\cal C}_{kl}$ will also lie
on a plane at a  distance from the origin and the  toric diagram 
on the plane for ${\cal C}_{23}$ is shown in
Figure 1.
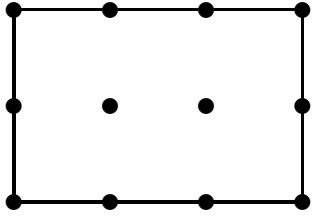
\begin{figure}
\setlength{\unitlength}{0.00083300in}%
\begingroup\makeatletter\ifx\SetFigFont\undefined%
\gdef\SetFigFont#1#2#3#4#5{%
  \reset@font\fontsize{#1}{#2pt}%
  \fontfamily{#3}\fontseries{#4}\fontshape{#5}%
  \selectfont}%
\fi\endgroup%
\begin{picture}(1366,1366)(500,-3444)
\thicklines
\put(3301,-2161){\circle*{100}}
\put(3301,-2761){\circle*{100}}
\put(3301,-3361){\circle*{100}}
\put(3901,-2161){\circle*{100}}
\put(3901,-2761){\circle*{100}}
\put(3901,-3361){\circle*{100}}
\put(4501,-3361){\circle*{100}}
\put(4501,-2761){\circle*{100}}
\put(4501,-2161){\circle*{100}}
\put(5101,-2161){\circle*{100}}
\put(5101,-2761){\circle*{100}}
\put(5101,-3361){\circle*{100}}
\put(3301,-2161){\line( 0,-1){1200}}
\put(3301,-3361){\line( 1, 0){1800}}
\put(5101,-3361){\line( 0, 1){1200}}
\put(5101,-2161){\line(-1, 0){1800}}
\end{picture}
\caption{A toric diagram for ${\bf Z}_2\times {\bf Z}_3$ hyper-quotient 
of the conifold, ${\cal C}_{23}$}
\end{figure}
In suitable  coordinates, the orbifolded conifold will be given by 
\bea
\label{con-eqn}
{\cal C}_{kl}: xy =z^l, \quad uv =z^k.
\eea

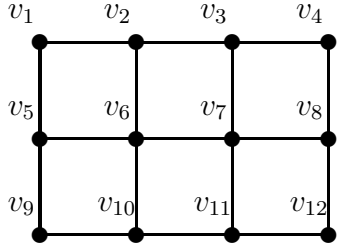
\begin{figure}
\setlength{\unitlength}{0.00083300in}%
\begingroup\makeatletter\ifx\SetFigFont\undefined%
\gdef\SetFigFont#1#2#3#4#5{%
  \reset@font\fontsize{#1}{#2pt}%
  \fontfamily{#3}\fontseries{#4}\fontshape{#5}%
  \selectfont}%
\fi\endgroup%
\begin{picture}(1366,2066)(500,-3444)
\thicklines
\put(3301,-2161){\circle*{100}}
\put(3301,-2761){\circle*{100}}
\put(3301,-3361){\circle*{100}}
\put(3901,-2161){\circle*{100}}
\put(3901,-2761){\circle*{100}}
\put(3901,-3361){\circle*{100}}
\put(4501,-3361){\circle*{100}}
\put(4501,-2761){\circle*{100}}
\put(4501,-2161){\circle*{100}}
\put(5101,-2161){\circle*{100}}
\put(5101,-2761){\circle*{100}}
\put(5101,-3361){\circle*{100}}
\put(3301,-2161){\line( 0,-1){1200}}
\put(3301,-3361){\line( 1, 0){1200}}
\put(4501,-3361){\line( 0, 1){1200}}
\put(4501,-2161){\line(-1, 0){1200}}
\put(3301,-2761){\line( 1, 0){1200}}
\put(3901,-2161){\line( 0,-1){1200}}
\put(5026,-2161){\line( 0, 1){  0}}
\put(4576,-2161){\line( 0, 1){  0}}
\put(4576,-2161){\line( 0, 1){  0}}
\put(4501,-2161){\line( 1, 0){600}}
\put(5101,-2161){\line( 0,-1){1200}}
\put(5101,-3361){\line(-1, 0){600}}
\put(4501,-2761){\line( 1, 0){600}}
\put(3100,-2011){$v_1$}
\put(3700,-2011){$v_2$}
\put(4300,-2011){$v_3$}
\put(4900,-2011){$v_4$}
\put(3100,-2611){$v_5$}
\put(3700,-2611){$v_6$}
\put(4300,-2611){$v_7$}
\put(4900,-2611){$v_8$}
\put(3100,-3211){$v_9$}
\put(3660,-3211){$v_{10}$}
\put(4260,-3211){$v_{11}$}
\put(4860,-3211){$v_{12}$}
\end{picture}
\caption{A toric diagram for $\widetilde{{\cal C}_{23}}$}
\label{partres}
\end{figure}

As we have seen above, the horizon   $X_5$ is singular.
To obtain a smooth Einstein manifold from  $X_5$, we  will resolve 
 the singularities of ${\cal C}_{kl}$ itself. We  resolve the singular threefold ${\cal C}_{kl}$ in  
two steps. In the first step, 
we choose a partial resolution, denoted by $\widetilde{{\cal C}_{kl}}$,
 of the orbifolded conifold ${\cal C}_{kl}$
for which the horizon will be smooth, but the Calabi-Yau threefold
$\widetilde{{\cal C}_{kl}}$  will have $kl$ number of isolated 
singular points.  Around each singular point, the Calabi-Yau space 
$\widetilde{{\cal C}_{kl}}$
is locally a metric cone over an Einstein manifold ${\T11}$.
In terms of the toric diagram, the partial resolution we have chosen
is obtained by adding all possible vertical and horizontal arrows to
the toric diagram of ${\cal C}_{kl}$.  
For example, the toric diagram for $\widetilde{{\cal C}_{23}}$
is given as in Figure \ref{partres}.

\begin{figure}
\centering
\epsfxsize=2in   
\hspace*{0in}\vspace*{.2in}
\epsffile{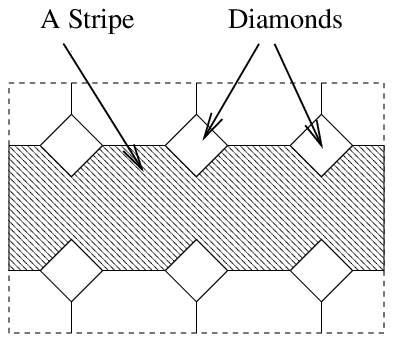}
\caption{A Brane Box Model} 
\label{branebox} 
\end{figure}

We are going to describe in detail each step but let us discuss first some features and make the connection of the
result with the T-dual brane configurations. The partially resolved space $\widetilde{{\cal C}_{kl}}$ is covered by
$kl$ squares and each square in the toric diagram represents an ordinary conifold. 
Thus the metric near each singular point can be written locally as follows
\bea
ds^2 = \frac{1}{9} (d\phi + \cos \theta_1 d\phi_1 + \cos \theta_2 d\phi_2)^2
+ \frac{1}{6} \sum_{a=1}^{2} (d\theta_a^2 + \sin^2 \theta_a d\phi^2_a).
\eea
Note that ${\cal C}_{kl}$ can be regarded as a ${\bf C^*}\times {\bf C^*}$
fibration over the $z$-plane via (\ref{con-eqn}).
 By taking  $T$-duality along $U(1) \times U(1)$
orbit in ${\bf C^*}\times {\bf C^*}$, we will have two types of NS branes
extended in, say, $x^0x^1x^2x^3x^4x^5$ and $x^0x^1x^2x^3x^8x^9$
 directions, where $x^4, x^8$ are compact 
directions coming from the degenerate $U(1) \times U(1)$-orbit. 
The separation of  the NS branes along the $x^8$ direction and
similarly that of the  NS' branes along the $x^4$ 
direction can be achieved by partially resolving ${\cal C}_{kl}$. But there 
will be $kl$ intersections of the NS and NS' branes on the $x^4x^8$ torus which correspond to the singular points of the 
partially resolved orbifolded conifold $\widetilde{{\cal C}_{kl}}$. 
By replacing these singular points by ${\P1}$'s, we  resolve the
singularities of $\widetilde{{\cal C}_{kl}}$. In the field theory, the process of blowing-up and turning the NS-NS
B-fluxes through the $kl$ ${\P1}$ cycles means turning on gauge couplings. In brane box configurations it means  
replacing the intersection of NS and NS' branes with `diamonds'~\cite{karch}, the size of the diamonds 
being given by the fluxes of $NSNS$ fields through the blow-up cycles. In this T-duality,
D3 branes become D5 branes which fill the $x^4x^8$ directions and the resulting brane configuration is as a brane
box model shown in Figure \ref{branebox}~\cite{hu,karch}. Its components are diamonds and boxes and a stripe is
an horizontal or vertical line of boxes (in Figure \ref{branebox} we have represented an horizontal stripe).
%%%%%%%%%%%%%%%%%%%%%%%%
%%%%%%%%%%%%%%%%%%%%%%%%%
In the second step we completely resolve the singularities of $\widetilde{{\cal C}_{kl}}$
by replacing each of the singular points by a copy of ${\P1}$ 
as explained before, procedure called a small resolution. In terms of the toric diagram, this 
corresponds to joining a pair of the diagonal vertices (but not both
pairs) by a line segment  in each square. Let us denote this completely
resolved threefold by $\widehat{{\cal C}_{kl}}$. 
For example, the toric diagram for $\widehat{{\cal C}_{23}}$
is given as in Figure 3.

\begin{figure}
\setlength{\unitlength}{0.00083300in}%
\begingroup\makeatletter\ifx\SetFigFont\undefined%
\gdef\SetFigFont#1#2#3#4#5{%
  \reset@font\fontsize{#1}{#2pt}%
  \fontfamily{#3}\fontseries{#4}\fontshape{#5}%
  \selectfont}%
\fi\endgroup%
\begin{picture}(1366,2066)(500,-3444)
\thicklines
\put(3301,-2161){\circle*{100}}
\put(3301,-2761){\circle*{100}}
\put(3301,-3361){\circle*{100}}
\put(3901,-2161){\circle*{100}}
\put(3901,-2761){\circle*{100}}
\put(3901,-3361){\circle*{100}}
\put(4501,-3361){\circle*{100}}
\put(4501,-2761){\circle*{100}}
\put(4501,-2161){\circle*{100}}
\put(5101,-2161){\circle*{100}}
\put(5101,-2761){\circle*{100}}
\put(5101,-3361){\circle*{100}}
\put(3301,-2161){\line( 0,-1){1200}}
\put(3301,-3361){\line( 1, 0){1200}}
\put(4501,-3361){\line( 0, 1){1200}}
\put(4501,-2161){\line(-1, 0){1200}}
\put(3301,-2761){\line( 1, 0){1200}}
\put(3901,-2161){\line( 0,-1){1200}}
\put(5026,-2161){\line( 0, 1){  0}}
\put(4576,-2161){\line( 0, 1){  0}}
\put(4576,-2161){\line( 0, 1){  0}}
\put(4501,-2161){\line( 1, 0){600}}
\put(5101,-2161){\line( 0,-1){1200}}
\put(5101,-3361){\line(-1, 0){600}}
\put(4501,-2761){\line( 1, 0){600}}
\put(3301,-2161){\line( 1,-1){1200}}
\put(4501,-3361){\line( 1, 1){600}}
\put(5101,-2761){\line(-1, 1){600}}
\put(4501,-2161){\line(-1,-1){1200}}
\put(3100,-2011){$v_1$}
\put(3700,-2011){$v_2$}
\put(4300,-2011){$v_3$}
\put(4900,-2011){$v_4$}
\put(3100,-2611){$v_5$}
\put(3700,-2611){$v_6$}
\put(4300,-2611){$v_7$}
\put(4900,-2611){$v_8$}
\put(3100,-3211){$v_9$}
\put(3660,-3211){$v_{10}$}
\put(4260,-3211){$v_{11}$}
\put(4860,-3211){$v_{12}$}
\end{picture}
\caption{A toric diagram for $\widehat{{\cal C}_{23}}$}
\end{figure}
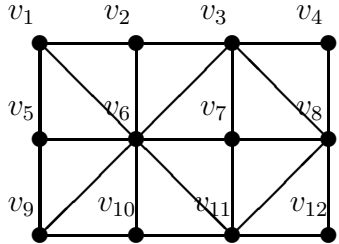

Let
\bea
\widetilde{\pi}: \widetilde{{\cal C}_{kl}} \to {\cal C}_{kl}, \quad
\widehat{\pi}: \widehat{{\cal C}_{kl}} \to \widetilde{{\cal C}_{kl}}
\label{resolution}
\eea
be the first and the second resolution discussed above. 
Let ${\bf o}$ be
the apex of the cone ${\cal C}_{kl}$ and
\bea
\widetilde{X_5} = \widetilde{\pi}^{-1}(X_5).
\eea
Note that $\widetilde{{\cal C}_{kl}}$
is covered by a $kl$ number of the ordinary conifolds corresponding
to the squares in the toric diagram. Hence $\widetilde{{\cal C}_{kl}}$
has $kl$ isolated singular points corresponding to the apexes of these
ordinary conifolds. These singular points lie on the fiber 
$\widetilde{\pi}^{-1}({\bf o})$. Moreover, we will explicitly show that
the exceptional fiber $\widetilde{\pi}^{-1}({\bf o})$ consists of $(k-1)(l-1)$ copies of ${\P1}\times {\P1}$.
The map  $\widehat{\pi}$ 
modifies only  the singular points of $ \widetilde{{\cal C}_{kl}}$ replacing
each of them by a copy of the projective space ${\P1}$. Thus
$\widehat{\pi}$ is an isomorphism outside
$(\widehat{\pi} \circ \widetilde{\pi})^{-1} ({\bf o})$. In particular,
we have
\bea
\widehat{\pi}^{-1}(\widetilde{X_5}) \cong \widetilde{X_5}.
\eea
and $\widetilde{X_5}$ is smooth.
As we mentioned above, $\widetilde{{\cal C}_{kl}}$ is smooth outside the 
 $kl$ ordinary conifold singular points.
Thus if we put a large number of D3 branes at one of $kl$ isolated 
singular points, denoted by $x_{ij}, 1\leq i \leq k, 1\leq j \leq l$,
then the near-horizon limit of the geometry will be
${\AdS5} \times {\T11}$.

The 5 dimensional manifold $\widetilde{X_5}$  
can be regarded as a smoothing of the singular Einstein manifold
$X_5$. As we will see later,
there will be $kl +k +l-2$ number of
2-cyles and 3-cycles in $\widetilde{X_5}$ where $kl$ is the number of the cycle coming 
from the horizon of each singular point of 
$ \widetilde{{\cal C}_{kl}}$ and $k+l -2$ number of them comes by
separating the above discussed two fixed circles in $X_5= \T11/\Z_k \times \Z_l$.
As mentioned above, the first kind of these cycles  corresponds  to the `diamonds' and
 the second kind
corresponds to the `stripes' in the brane box model. They correspond to stripes of boxes instead of
individual boxes because we need to consider curves of either ${\A_{k-1}}$ or ${\A_{l-1}}$ 
singularity.

Before starting the actual discussion concerning the identification of the different 2-cycles, 
we present another proof for the fact that  the Einstein manifold ${\bf T}^{1,1}$ is homeomorphic to 
${\bf S}^3 \times {\bf S}^2$~\cite{kw, co}.
By changing coordinates
\bea
z_1  = w_1 + i w_2,\quad
z_2 = w_1 - i w_2,\quad z_3 = w_3 + i w_4,\quad z_4 = w_3 - i w_4,
\eea
we can rewrite the conifold equation (\ref{conieqn}) as:
\bea
\label{conieqn2}
w_1^2 + w_2^2 + w_3^2 + w_4^2 = 0.
\eea
Since the Einstein manifold ${\bf T}^{1,1}$ can be realized as a horizon (link)
of the the conifold singularity, ${\bf T}^{1,1}$ is described by the 
intersection of (\ref{conieqn2}) and the seven sphere in $\C^4$ given by
\bea
\label{seven}
|w_1|^2 + |w_2|^2 +|w_3|^2 + |w_4|^2 = 1.
\eea
>From (\ref{conieqn2}) and (\ref{seven}), we see that ${\bf T}^{1,1}$ 
is given by
\bea
\label{t11}
x_1^2 + x_2^2 + x_3^2 + x_4^2 = y_1^2 + y_2^2 + y_3^2 + y_4^2 = 1/2,
\nonumber \\
x_1y_1 + x_2y_2 + x_3y_3 + x_4y_4 =0,
\eea
where $x_i$ and $y_i$ are the real and imaginary parts of $w_i$.
Thus the $y_i$'s describe a bundle of two spheres  in the tangent bundle
of ${\bf S^3}$ given by the coordinates $x_i$'s.
Hence ${\bf T}^{1,1}$ is a sphere bundle ${\bf S^2}$ over ${\bf S^3}$.
Since ${\bf S^3}$ is parallelizable~\cite{steenrod}, 
a sphere bundle ${\bf S^2}$ over ${\bf S^3}$ is trivial and
${\bf T}^{1,1}$ is diffeomorphic to ${\bf S^2} \times {\bf S^3}$.
In fact, the frame for the sphere bundle over ${\bf S^3}$ can be given by
\bea
\{ (x_2, -x_1, -x_4, x_3),\quad ( x_3, x_4, -x_1, -x_2),\quad
 (x_4, -x_3, x_2, -x_1) \}.
\eea

Next, we 
want to study the exceptional fiber $\widetilde{\pi}^{-1}({\bf o})$.
We will illustrate the general situation with $\widetilde{{\cal C}_{23}}$.
To facilitate understanding, let us choose a basis for the lattice $\N$
so that the coordinates of the lattice points are as follows:
\bea 
\begin{array}{llll}
v_1 = (1,0,1),\quad &v_2 = (1,1,1),\quad &v_3 = (1,2,1),\quad 
                                                   &v_4 = (1,3,1),\nonumber \\ 
v_5 =(1,0,0),\quad &v_6= (1,1,0),\quad &v_7= (1,2,0),  \quad
                                                   &v_8= (1,3,0), \nonumber \\  
v_9 = (1,0,-1), \quad &v_{10} =(1,1,-1),\quad &v_{11} =(1,2,-1).
                                           \quad &v_{12} =(1,3,-1).
\end{array}
\eea
The Figure \ref{coord} shows the coordinate rings corresponding to
the various squares of the toric diagram of $\widetilde{{\cal C}_{23}}$.
\begin{figure}

\[
\begin{array}{|c|c|c|}
\hline
\begin{tabular}{c}The vertices of\\the square\end{tabular}   & 
\begin{tabular}{c}The coordinate rings\\ 
\begin{tabular}{c}The ideal of \\
the fiber $\widetilde{\pi}^{-1}({\bf o})$\end{tabular}\end{tabular}&
\begin{tabular}{c}The restriction of  \\the map $\widetilde{\pi}$\end{tabular} 
                        \\ \hline
\{ v_1, v_2, v_5, v_6\}   &
    \begin{array}{c} \C [z, y, xy^{-1}, xz^{-1}]\\ 
               (y,xz^{-1}) \end{array}
   &  \begin{array}{llll}
          a = z,& b=y,& c= xy^{-1}, &d= xz^{-1}\\ 
          p = b,& q=b^2c^3,& r = a^2d,& s=d
    \end{array}  \\ \hline
\{ v_5, v_6, v_9, v_{10} \}&
     \begin{array}{c} \C [z^{-1}, y, xy^{-1}, xz]\\ (y, xz)\end{array}
      & \begin{array}{llll}
          a = z^{-1},& b=y,& c= xy^{-1}, &d= xz\\ 
          p = b,& q=b^2c^3,& r =d ,& s=a^2d
    \end{array} \\ \hline
\{ v_2, v_3, v_6, v_{7} \} &\begin{array}{c}
                           \C [z, x^2y^{-1}, x^{-1}y, xz^{-1}]\\
                             (xz^{-1})\end{array}
     & \begin{array}{llll}
          a = z,& b=x^2y^{-1},& c= x^{-1}y, &d= xz^{-1}\\ 
          p = bc^2,& q=b^2c,& r =a^2d ,& s=d
    \end{array} \\ \hline
\{  v_6, v_7, v_{10}, v_{11} \}&\begin{array}{c}
\C [ z^{-1}, x^2y^{-1},x^{-1}y,xz]\\
(xz)\end{array}&\begin{array}{llll}
          a = z^{-1},& b=x^2y^{-1},&c=x^{-1}y,& d=xz\\ 
          p = bc^2,& q=b^2c,& r =d ,& s=a^2d
    \end{array}
 \\ \hline
\{  v_3, v_4, v_{7}, v_{8} \}&\begin{array}{c}
\C [z,  x^3y^{-1},  x^{-2}y, xz^{-1}]\\
(x^3y^{-1},xz^{-1})\end{array}&\begin{array}{llll}
          a = z,& b=x^3y^{-1},& c=x^{-2}y,& d=xz^{-1}\\ 
          p = b^2c^3,& q=b,& r =a^2d ,& s=d
    \end{array}
\\ \hline
\{  v_7, v_8, v_{11}, v_{12} \}& \begin{array}{c}
\C [ z^{-1}, x^3y^{-1},x^{-2}y, xz]\\
( x^3y^{-1},xz)\end{array}& \begin{array}{llll}
          a = z^{-1},& b=x^3y^{-1},& c=x^{-2}y,& d=xz\\ 
          p = b^2c^3,& q=b,& r =d ,& s=a^2d
    \end{array}
\\ \hline
\{  v_1, v_4, v_9, v_{12} \} &  
\C [y, x^3y^{-1}, xz, xz^{-1}]
&(y,x^3y^{-1},xz,xz^{-1})\\& &  \\
 \hline
\end{array}
\]

\caption{The coordinate rings for the squares}
\label{coord}
\end{figure}
By tedious but  direct computations from  the Figure \ref{coord}, 
one can see that
the fiber $\widetilde{\pi}^{-1}({\bf o})$ consists of a union of $(k-1)(l-1)$
numbers of ${\P1}\times {\P1}$. Moreover the adjacent components
of the fiber meet along ${\P1}$.
 If we denote each 
${\P1}\times {\P1}$ by a vertex  and we join two of vertices if
they meet, then we get a lattice of size $(k-1)$ by $(l-1)$.
The Figure \ref{excc23} shows what they look like for 
$\widetilde{{\cal C}_{23}}$.
\begin{figure}
\label{excc23}
\setlength{\unitlength}{0.000483300in}%
\begingroup\makeatletter\ifx\SetFigFont\undefined%
\gdef\SetFigFont#1#2#3#4#5{%
  \reset@font\fontsize{#1}{#2pt}%
  \fontfamily{#3}\fontseries{#4}\fontshape{#5}%
  \selectfont}%
\fi\endgroup%
\begin{picture}(4966,4895)(-1000,-5244)
\thicklines
\put(2401,-2161){\circle*{150}}
\put(2401,-5161){\circle*{150}}
\put(4801,-961){\circle*{150}}
\put(4801,-3961){\circle*{150}}
\put(7201,-2161){\circle*{150}}
\put(7201,-5161){\circle*{150}}
\put(2401,-2161){\line( 0,-1){3000}}
\put(2401,-5161){\line( 2, 1){3600}}
\put(6001,-3361){\line( 0, 1){3000}}
\put(6001,-361){\line(-2,-1){3600}}
\put(3601,-361){\line( 2,-1){3600}}
\put(7201,-2161){\line( 0,-1){3000}}
\put(7201,-5161){\line(-2, 1){3600}}
\put(3601,-3361){\line( 0, 1){3000}}
\put(4801,-961){\line( 0, 1){  0}}
\put(4801,-961){\line( 0, 1){  0}}
\multiput(4801,-961)(0.00000,-9.00901){334}{\line( 0, -1){1}}
\end{picture}
\caption{The fiber $\widetilde{\pi}^{-1}({\bf o})$ for 
$\widetilde{{\cal C}_{23}}$}
\end{figure}
In Figure \ref{excc23},  each square represents ${\P1}\times {\P1}$ and
the singular points of $\widetilde{{\cal C}_{23}}$ are denoted by black
dots.
The fiber
 consists of two ${\P1}\times {\P1}$ and they meet along ${\P1}$.
>From this picture, one can see that the second betti number 
$h_2(\widetilde{\pi}^{-1}({\bf o})) = k+l -2$. The D5 brane
wrapping one of these $k+l -2$ spheres corresponds  the
fractional D3 brane coming from the ${\bf Z}_k \times {\bf Z}_l$
twisted sector of the type IIB supergravity on  ${\AdS5}\times 
{\T11}/{\bf Z}_k \times {\bf Z}_l$. In the brane box model,
this type of a fractional D3 brane
turns into a D5 brane living on a stripe (See Figure \ref{branebox}).

As we mentioned above, a full resolution $\widehat{{\cal C}_{kl}}$
of ${\cal C}_{kl}$ can be obtained by replacing the singular points of
$\widetilde{{\cal C}_{kl}}$ by copies of ${\bf P}^1$.
Hence the exceptional fiber 
$(\widehat{\pi} \circ \widetilde{\pi})^{-1} ({\bf o})$ will acquire 
$kl$ copies of ${\P1}$. This can be achieved by blowing up
$kl$ points on $\widetilde{{\pi}^{-1}} ({\bf o})$. Thus the second betti
number of the exceptional fiber 
$(\widehat{\pi} \circ \widetilde{\pi})^{-1} ({\bf o})$ will be
$k+l -2 +kl$.

To study cycles on the Einstein manifold $\widetilde{X_5}$, consider 
an inclusion  map
\bea
i: \widetilde{X_5} \hookrightarrow \widehat{{\cal C}_{kl}}.
\eea
First note that $\widehat{{\cal C}_{kl}}$ contacts to the
exceptional fiber $(\widehat{\pi} \circ \widetilde{\pi})^{-1} ({\bf o})$, which
we will denote by $E$.
This contraction can be constructed by lifting a conical structure of
${\cal C}_{kl}$. 
We now regard $\widetilde{X_5}$ as a boundary of a smooth 6 dimensional
manifold $\widehat{{\cal C}_{kl}}$. Consider a 
long sequence of homology groups with ${\bf Q}$ coefficients:
\bea
H_4(\widehat{{\cal C}_{kl}}, \widetilde{X_5}) 
\stackrel{\partial}{\longrightarrow} 
H_3(\widetilde{X_5}) \stackrel{i_\ast}{\longrightarrow}
 H_3(\widehat{{\cal C}_{kl}})
\stackrel{j_\ast}{\longrightarrow}
H_3(\widehat{{\cal C}_{kl}}, \widetilde{X_5}) 
\stackrel{\partial}{\longrightarrow} 
H_2(\widetilde{X_5})\stackrel{i_\ast}{ \longrightarrow}  H_2(\widehat{{\cal C}_{kl}}) 
\eea
where $j_\ast$ is  induced by the inclusion 
$j: \widehat{{\cal C}_{kl}} \subset 
(\widehat{{\cal C}_{kl}}, \widetilde{X_5})$.  Via the universal coefficient
theorem and the Poincar\'e duality, we have
\bea
H_3(\widehat{{\cal C}_{kl}}, \widetilde{X_5}) 
\cong  H^3(\widehat{{\cal C}_{kl}}) \cong H_3(\widehat{{\cal C}_{kl}}) = 0
\eea 
since $\widehat{{\cal C}_{kl}}$ can be deformed to $E$.
Hence in the following commutative diagram, the top horizontal arrow will be
injective and the bottom horizontal arrow will be surjective.
\\
\setlength{\unitlength}{0.05mm}
\begin{picture}(1000, 1200)(-600,-80)
\put (100, 800){$H_2(\widetilde{X_5}, {\bf Q})$}
\put (100, 500){$H^3(\widetilde{X_5}, {\bf Q})$}
\put (100, 200){$H_3(\widetilde{X_5}, {\bf Q})$}
\put (1100, 800){$H_2(\widehat{{\cal C}_{kl}}, {\bf Q})$}
\put (1100, 500){$H^2(\widehat{{\cal C}_{kl}}, {\bf Q})$}
\put (1050, 200){$H_4(\widehat{{\cal C}_{kl}},\widetilde{X_5}, {\bf Q})$}
\put (550, 820){\vector(1,0){500}}
\put (1000, 220){\vector(-1,0){450}}
\put (280, 750){\vector(0,-1){130}}
\put (250, 680){$\wr$}
\put (280, 450){\vector(0,-1){130}}
\put (250, 380){$\wr$}
\put (1280, 750){\vector(0,-1){130}}
\put (1250, 680){$\wr$}
\put (1280, 450){\vector(0,-1){130}}
\put (1250, 380){$\wr$}
\put (750, 840){$i_\ast$}
\put (750, 240){$\partial$}
\end{picture}
\\
Here the vertical arrows are isomorphisms because of the universal coefficient
theorem and the Poincar\'e dualities.
Therefore, we conclude that
\bea
\label{2-cycle}
H_2(\widetilde{X_5}, {\bf Q}) \cong H_2(\widehat{{\cal C}_{kl}}, {\bf Q}).
\eea

Moreover, we want to see the origin of these two cycles.
Let $L_{ij}$ be the horizon of
$\tCkl$ at $x_{ij}$. Then $L_{ij}$ is isomorphic to $\T11$ and
$L_{ij}$ does not change under the resolution 
$\widehat{\pi}: \widehat{{\cal C}_{kl}} 
\to \tCkl$. From (\ref{2-cycle}), we see that
there is a natural inclusion 
\bea
H_2(L_{ij}) \hookrightarrow H_2(\hCkl) \cong H_2(\tX5).
\eea
Thus we may regard the 2-cycle of $H_2(L_{ij})$ as a 2-cycle of $H_2(\tX5)$.
We denote this 2-cycle by $C^2_{ij}$.
By Poincar\'e duality, we may also regard the 3-cycle of $H_3(L_{ij})$
as a 3-cycle of $H_3(\tX5)$, which will be denoted by $C^3_{ij}$.
Note that there are $kl$ contributions of 2-cycles from $H_2(L_{ij})$.

Moreover on each open neighborhood of $x_{ij}$ represented by
a square, we can choose a basis for one-forms
\bea
 e_{ij}^\psi = \tf{1}{3} \left( d\psi + \cos\theta_1 \phi_1 + 
    \cos\theta_2 \phi_2 \right)  \cr
   e_{ij}^{\theta_1} = \tf{1}{\sqrt{6}} d\theta_1 \qquad
   e_{ij}^{\phi_1} = \tf{1}{\sqrt{6}} \sin\theta_1 d\phi_1  \cr
   e_{ij}^{\theta_2} = \tf{1}{\sqrt{6}} d\theta_2 \qquad
   e_{ij}^{\phi_2} = \tf{1}{\sqrt{6}} \sin\theta_2 d\phi_2 \ ,
\eea
 so that the harmonic representatives of the second and third
cohomology groups can be written as 
\bea
   e_{ij}^{\theta_1} \wedge e_{ij}^{\phi_1} - e_{ij}^{\theta_2} 
\wedge e_{ij}^{\phi_2}
    &\in H^2(L_{ij} )
  \cr
   e_{ij}^\psi \wedge e_{ij}^{\theta_1} \wedge e_{ij}^{\phi_1} - 
    e_{ij}^\psi \wedge e_{ij}^{\theta_2} \wedge e_{ij}^{\phi_2} &\in 
H^3(L_{ij}) \ .
\eea

On the other hand, from the inclusion of $\tX5$ into the partially resolved 
conifold $\tCkl$, we obtain a map
\bea
H_2(\tX5) \rightarrow H_2(\tCkl).
\eea
Since we obtain $\tCkl$ from $\hCkl$ by collapsing the blown-up
two spheres which is a smooth deformation of  the spheres in
$L_{ij}$, we have the following exact sequence:
\bea
0\rightarrow \bigoplus_{i=1,\ldots ,k, j=1,\ldots ,l}
H_2(L_{ij}) \rightarrow H_2(\tX5)\rightarrow H_2(\tCkl)
\rightarrow 0.
\eea
Therefore we have  obtained a concrete description of the cycles of $\tX5$
in terms of $kl$ cycles  from $H_2(L_{ij})$
and the $k+l-2$ cycles $H_2(\tCkl)$.
The cycles in $H_2(\tCkl)$ are separating the singular points of
$\tCkl$. hence the corresponding cycles in $H_2(\tX5)$ will separate
the two fixed circles of $X_5$.
 This  generalizes the results from the conifold and allows 
us to study the fluxes of NS-NS and R-R two forms in order to
obtain logarithmic  RG flow in the next section.

%%%%%%%%%%%%%%%%%%%
%%%%%%%%%%%%%%%%%%%%%%
\section{Fractional Branes and RG Flows}
\setcounter{equation}{0}
In this section we are going to extensively use the mathematical results of the previous section and
identify the fractional branes as small perturbations of the string background.
This will allow us to study the interpolation between the background with or without
fractional D3 branes.
This description will be shown to reproduce the logarithmic flow of gauge couplings,being
in complete agreement with results of field theory.

We begin with a brief review of \cite{kn} where the RG flow 
determined by the fractional D3 branes was considered. Their result 
is a particular example of our case for $k=l=1$.  
The coupling constant of field theory are written in terms of the two-form charges
 on the vanishing sphere of the singularity:
\begin{equation}
\tau = C_0 + i \frac{1}{g^2} = \int_{C^2} B^{RR} + i \int_{C^2} B^{NS}
\end{equation}
where $B^{RR}, B^{NS}$ are the R-R and NS-NS 2-form potentials. 
At the conifold point the values of the B-fields are fixed
and the coupling constant is $g^{-2} \sim e^{-\phi}/2$.
By wrapping $M$ D5 branes over the 2-cycle of $\T11$ in addition to
$N$ regular D3 branes orthogonal to the conifold,  
the string background will contain 
$M$ units of RR 3-form flux through the 3-cycle of 
${\bf T}^{1,1}$:
\begin{equation}
\int_{C^3} H^{RR} = M.
\end{equation}
The equations of motion imply that $H^{NS}$ should be proportional to $M$ and to
a product of 
the closed 2-form on ${\bf T}^{1,1}$ and a one form which involves $dr$ and taken to be
  $\frac{df}{dr} dr = d f(r)$.
Hence the two form potential 
$B^{NS}$ will be:
\begin{equation}
\label{nsns1}
B^{NS} = e^{\phi} f(r) \omega_2, \quad \mbox{where}\quad
 f(r) \sim M \log r
\end{equation}
where $\omega_2$ is the closed form on ${\bf T}^{1,1}$.
The R-R scalar $C_0 =0$ and the dilaton are set to have constant values.

The fractional D3 branes, obtained by wrapping D5 branes on the 2-cycle of  ${\bf T}^{1,1}$
represent domain walls in ${\AdS5}$ and are obtained by wrapping D5 branes on 
the 2-cycle of  ${\bf T}^{1,1}$. 
The relation between the two coupling constants  of  the field theory on the D3 branes writes in
the presence of M  D5 branes wrapped on the 2-cycle as 
\begin{equation}
\frac{1}{g_1^2} - \frac{1}{g_2^2} \sim e^{- \phi} (M \log r -
\frac{1}{2}) \sim e^{-\phi} M \log r
\end{equation}
where the last relation is true in the large M approximation.
This agrees  with the field theory logarithmic RG flow equation in a non-conformal theory, the
conformality being broken by the presence of the fractional D3 branes.

We now proceed to the case of the orbifolded conifold.
The field theory on the world-volume of the $N$ coincident D3 branes probing the singularity
${\cal C}_{kl}$ has been obtained in \cite{ura}. 
It  is an ${\cal N} = 1$ chiral supersymmetric gauge theory with the gauge group 
\bea
\prod_{i=1}^k \prod_{j=1}^l SU(N)_{i,j} \times \prod_{i=1}^k \prod_{j=1}^l 
SU(N)'_{I,j}
\eea
and with  matter fields 
\begin{center}
\begin{tabular}{ll}
{\bf Field} & {\bf Representation} \\
$(A_1)_{i+1,j+1;i,j}$ & $(\fund_{i+1,j+1},\antifund\; '_{i,j})$ \\
$(A_2)_{i,j;i,j}$ & $(\fund_{i,j},\antifund\; '_{i,j})$ \\
$(B_1)_{i,j;i,j+1}$ & $(\fund\; '_{i,j},\antifund_{i,j+1})$ \\
$(B_2)_{i,j;i+1,j}$ & $(\fund\; '_{i,j},\antifund_{i+1,j})$
\end{tabular}
\end{center}
Moreover there is a quartic superpotential 
\bea
\label{superpot}
W \sim \sum \left((A_1)_{i+1,j+1;i,j} (B_1)_{i,j;i,j+1} 
(A_2)_{i,j+1;i,j+1} (B_2)_{i,j+1;i+1,j+1} -\right. \\ \nonumber
\left. (A_1)_{i+1,j+1;i,j} (B_1)_{i,j;i+1,j} (A_2)_{i+1,j;i+1,j} \
(B_2)_{i+1,j;i+1,j+1}\right)
\eea
As explained in the section  2, by taking T-duality, we obtain
the brane box configuration consisting  of $k$ NS branes and $l$ NS' branes 
whose intersections are smoothen out
by diamonds. The singular point of ${\cal C}_{kl}$
splits into $kl$ ordinary conifold singularities $x_{ij}$
 on $\tCkl$ under the resolution
$\widetilde{\pi}: \widetilde{{\cal C}_{kl}} 
\to {\cal C}_{kl}$ as in equation ~(\ref{resolution}).  Hence if we put a large $N$ number of
D3 branes at each singular point $x_{ij}$,  
 the the near-horizon limit of the geometry of $\tCkl$ at
$x_{ij}$  will be $\AdS5 \times \T11$. 
If we wrap a D5 brane over the 2-cycle of $\tX5$ corresponding
to the 2-cycle of $L_{ij}$, we obtain a fractional D3 brane which is 
a  domain wall in ${\AdS5}$ since it lies in the orthogonal direction
to the D3 branes placed on the singular point $x_{ij}$.
When we put one fractional brane together with $N$ regular D3 branes, we will change the $(i,j)$-th
copy of the $SU(N)$ gauge group and the gauge group
will change to  $SU(N+1) \times SU(N)^{kl-1} \times SU(N)^{'kl}$ on the other
side of the domain wall. The evidence for this claim is 
similar to the one of \cite{gk} i.e. by studying the behavior of 
wrapped D3-branes on 3-cycles of $L_{ij}$ when they cross domain walls.

Before going further,
we need to make a crucial observation concerning the constraint imposed by the consistency of the
field theory on the worldvolume of the D3 branes.  Since
our field theory is chiral and can have anomalies, it
is important to be careful with the way we introduce the fractional D3 branes. 
>From the geometrical discussion, it appears that there is no restriction on introducing fractional 
D3 branes i.e. on wrapping D5 branes on different 2-cycles of the horizon. 
In the brane box picture this 
would mean that there is no restriction on the number 
of D5 branes on different diamonds. 
If there is no integer D3 brane in the theory, 
we can introduce any number of fractional 
D3 branes which correspond to D5 branes in a specific diamond. 
In the presence of D3 branes orthogonal 
to the conifold (integer D3 branes), we cannot
put fractional D3 branes in only one diamond. 
This is because if we put one D5 brane in the
$(i,j)$-th diamond, the gauge groups 
in the $(i, j)$-th and $(i+1,j+1)$-th boxes have one supplementary 
anti-fundamental field and the gauge groups in the $(i,j-1)$-th and 
$(i-1,j)$-th boxes have one supplementary
fundamental field, these four gauge theories becoming anomalous. 
See Figure 8 where we represent the $ij$ diamond and fields which are in the fundamental or
anti-fundamental representations.
This determines a specific way to introduce fractional branes. 
We need to have
either $k$ fractional D3 branes corresponding to D5 branes on a row of diamonds or $l$ fractional
D3 branes corresponding to D5 branes on a column of diamonds. 
Then all the gauge groups in different boxes have the same number of fundamental and anti-fundamental 
fields and are anomaly free. 
When we discussed about the twisted sector in section 2, we saw that they also correspond to 
D5 branes on rows or columns of boxes so the filling of boxes and diamonds is similar.
%%%%%%%%%%%%%%%%%%%%%%%%
\begin{figure}
\centering
\epsfxsize=2in   
\hspace*{0in}\vspace*{.2in}
\epsffile{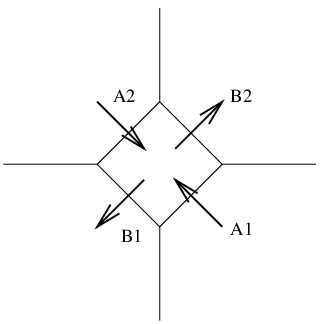}
\caption{A Diamond with the Corresponding Chiral Fields} 
\label{fields} 
\end{figure}
%%%%%%%%%%%%%%%%%%%%%%%%%

We can now proceed to obtain the main goal of this section i.e. to compare 
the $\beta$-function calculation in field theory living on the world-volume of the integer D3 branes
with the solution of supergravity equations of motion in the presence of fractional D3 branes.  
To start with, we need to discuss more about the ``dibaryon'' operators and the domain
walls in $AdS_5$. Because $\tX5$ contains $kl$ copies of Einstein manifolds ${\bf T}^{1,1}$, 
we can wrap D3 branes over the 3-cycle of each ${\bf T}^{1,1}$ to obtain $kl$ types of ``dibaryons''. 
Besides we have integer D3 branes  and 
D5 branes wrapped on each of the $kl$ blow-up 2-cycles at each of the $kl$ 
singular points, the latter ones being the domain walls. 
For $M_{i_{0}j_{0}}$ D5 branes wrapped around the $(i_{0},j_{0})$-th 2-cycle,
the gauge group changes from  $SU(N)^{kl} \times SU(N)^{'kl}$ to  
\begin{equation}
\label{ngroup}
SU(N + M_{i_{0}j_{0}}) \prod_{i=1}^{k}\prod_{j=1}^{l}  SU(N)_{ij} \times SU(N)^{'kl}
\end{equation}
where the pair $(i, j)$ does not take the value $(i_0, j_0)$.
As discussed before, field theory results require D5 branes on either rows or columns of diamonds. 
and this means that we need to have  $M_{i_{0}j},  j = 1,\cdots,l$ D5 branes wrapped 
around the $i_{0}, j=1,\cdots,l$ cycles, the gauge group changing to
\begin{equation}
\label{ngroup1}
 \prod_{j=1}^{l} SU(N + M_{i_{0} j}) \prod_{i=1, i \ne i_0}^{k} \prod_{j=1}^{l} SU(N)_{ij} \times SU(N)^{'kl}
\end{equation}
For  $M_{i j_{0}}, i = 1,\cdots,k$ D5 branes wrapped around the $i=1,\cdots,l; j_0$ cycles the gauge group changes to
\begin{equation}
\label{ngroup2}
\prod_{i=1}^{k} SU(N + M_{i j_{0}}) \prod_{i}^{k} 
\prod_{j=1, j \ne j_0}^{l} SU(N)_{ij} \times SU(N)^{'kl}
\end{equation}

We can now proceed to construct the Type IIB dual to the ${\cal N} = 1$ supersymmetric field
theory with the gauge group (\ref{ngroup1}) or (\ref{ngroup2}). 
We have $kl$ fluxes of RR 3-form through the 3-cycles $C_{ij}$ which are 
Hodge duals to the 2-cycles surrounding the singular points
$x_{ij}$ for $ i= 1, \ldots, k, j=1, \ldots, l$:
\begin{equation}
\int_{C^3_{ij}} H^{RR} = M_{ij},\,  i= 1, \ldots, k, j=1, \ldots, l
\end{equation}
Here we are identifying a 2-cycle of $H_2(L_{ij})$ with a 2-cycle of 
$H_2(\tX5)$. 
To obey the above observed rule, one needs to turn fluxes through all $C^{3}_{i_0 j}, j=1,\cdots,l$ or
all $C^3_{i j_0}, i=1,\cdots,k$ cycles with fluxes equal to  
$M_{i_0 j}$ or $M_{i j_0}$ respectively, $i_0, j_0$ being some fixed indices.

We can now use the results of our previous section where we have completely identified the 2-cycles and the 3-cycles
so the result is that the $H^{RR}$ which we need to turn on are:
\bea 
\label{HRR1}
 H^{RR} \sim 
\sum_{j} M_{i_0 j} 
 e_{i_0 j}^\psi \wedge ( e_{i_0 j}^{\theta_1} \wedge e_{i_0 j}^{\phi_1} - 
    e_{i_0 j}^{\theta_2} \wedge e_{i_0 j}^{\phi_2} ),\quad  \mbox{for fixed} \quad i_0 \quad  \mbox{and} \quad j = 1,\cdots,l
\eea 
or
\bea 
\label{HRR2}
 H^{RR} \sim 
\sum_{i} M_{i j_0} 
 e_{i j_0}^\psi \wedge ( e_{i j_0}^{\theta_1} \wedge e_{i j_0}^{\phi_1} - 
    e_{i j_0}^{\theta_2} \wedge e_{i j_0}^{\phi_2} ),\quad  \mbox{for fixed} \quad j_0 \quad \mbox{and} \quad i = 1,\cdots,k.
\eea 

We now  consider the Type IIB SUGRA equations of motion with the 2-form
gauge potentials in the ${\AdS5} \times X_5$ background
with constant $\tau = C_0 + i e^{-\phi}$:
\bea
\label{eom}
d\ast G = i F_5 \wedge G.
\eea
Here $G$ is the complex 3-form field strength,
\bea
G = H^{RR} + \tau H^{NSNS},
\eea
which satisfies the Bianchi identity $dG = 0$.

If we choose  $C_0 =0$ and a constant dilaton, it follows from (\ref{HRR1}), 
(\ref{HRR2}) and (\ref{eom}) that
\bea
e^{-\phi}H^{NSNS} \sim 
\sum_{j} df_{i_0 j}(r) \wedge ( e_{i_0 j}^{\theta_1} \wedge e_{i_0 j}^{\phi_1} - 
    e_{i_0 j}^{\theta_2} \wedge e_{i_0 j}^{\phi_2} ), \mbox{for fixed} \quad i_{0}
\eea
or 
\bea
e^{-\phi}H^{NSNS} \sim 
\sum_{i} df_{i j_0}(r) \wedge ( e_{i j_0}^{\theta_1} \wedge e_{i j_0}^{\phi_1} - 
    e_{i j_0}^{\theta_2} \wedge e_{i j_0}^{\phi_2} ), \mbox{for fixed} \quad j_{0}
\eea
are two solutions for the NS 3-form corresponding to the specific choice for
$H^{RR}$.
Since 
$F_5 = \mbox{vol}({\AdS5}) + \mbox{vol} (X_5)$ and 
$d\ast H^{RR} = -e^{-\phi} F_5 \wedge H^{NSNS}$ , we conclude
\bea
F_5 \wedge H^{NSNS} =0
\eea
and the real part of (\ref{eom}) is satisfied for all $f_{ij}$.
>From the imaginary part we have either
\bea
\frac{1}{r^3} \frac{d}{dr} \left( r^5 \frac{d}{dr} f_{i_0 j}(r) \right) \sim M_{i_0 j}, \mbox{for fixed} \quad i_{0}
\eea
or
\bea
\frac{1}{r^3} \frac{d}{dr} \left( r^5 \frac{d}{dr} f_{i j_0}(r) \right) \sim M_{i j_0}, \mbox{for fixed} \quad j_{0}
\eea
Thus we need to turn on an NS form as
\bea
B^{NSNS} \sim e^{\phi} \sum_{j} M_{i_0 j}\omega_{i_0 j} \log r , \mbox{for fixed} \quad i_{0}
\eea
or
\bea
B^{NSNS} \sim e^{\phi} \sum_{i} M_{i j_0}\omega_{i j_0} \log r , \mbox{for fixed} \quad j_{0}
\eea

The gauge couplings of the gauge theories are 
modified in the presence of the fluxes of the $B^{NS}$ through the various 2-cycles.
The gauge coupling without B-flux is related to the string coupling constant 
as $g^{-2} = \frac{1}{2 g_{s}}$  where $g_s$ is the string coupling constant.
If all the diamonds and boxes have the same area, then field theories corresponding to D5 branes on boxes and diamonds have the same
coupling constant and this the meaning of $g$ in the previous formula. 
Since the B-fields (inverse of the gauge couplings) are areas on the torus, by changing 
the B-fluxes through the $(i_0, j), j=1,\cdots,l$ or $(i ,j_0), i=1,\cdots,k$ cycles 
we change the areas of the diamonds. 
If we wrap D5 branes on the  $(i_0, j), j=1,\cdots,l$ or $(i, j_0), i=1,\cdots,k$ cycles, the fluxes of B-field through the
corresponding cycles modify and the gauge couplings change acoording to 
$\frac{1}{g_s}\int_{C^2_{i_0j}} B^{NSNS}, j=1,\cdots,l$ or $\frac{1}{g_s}\int_{C^2_{i j_0}} B^{NSNS}, i=1,\cdots,k$. 
The connection to the RG flow in field theory uses the relation: 
\begin{equation}
\label{rg1}
\frac{1}{g_{i_0^2 j}} - \frac{1}{g^2} \sim \frac{1}{g_s} (\int_{C^2_{i_0 j}} B^{NSNS} - 1/2), \mbox{for fixed} \quad i_{0} \quad 
\mbox{and} \quad j=1,\cdots,l
\end{equation}
or 
\begin{equation}
\label{rg2}
\frac{1}{g_{i j_0}^2} - \frac{1}{g^2} \sim \frac{1}{g_s} (\int_{C^2_{i j_0}} B^{NSNS} - 1/2),  \mbox{for fixed} \quad j_{0} \quad 
\mbox{and} \quad i=1,\cdots,k
\end{equation}
The previous results for $B^{NSNS}$ allow us to rewrite  (\ref{rg1}) as 
\begin{equation}
\frac{1}{g_{i_0 j}^2} - \frac{1}{g^2} \sim M_{i_0 j} \log r,\,\, \mbox{for fixed} \quad i_{0} \quad 
\mbox{and} \quad j=1,\cdots,l
\label{rg3}
\end{equation}
and (\ref{rg2}) can be written as
\begin{equation}
\frac{1}{g_{i j_0}^2} - \frac{1}{g^2} \sim M_{i j_0} \log r,\,\, \mbox{for fixed} \quad j_{0} \quad 
\mbox{and} \quad i=1,\cdots,k
\label{rg4}
\end{equation}
Because the coordinate $r$ is seen as a field theory scale in the AdS/CFT conjecture, relations (\ref{rg3}) and 
(\ref{rg4}) give the supergravity dual of the scale dependence of the difference between the gauge couplings.

The above results are obtained by solving the supergravity equations of motion  and we are now going to compare them 
with $\beta$-function calculations in ${\cal N}=1$ supersymmetric field theory which give:
\begin{eqnarray}
\label{rg11}
\frac{d}{d\,\log (\Lambda/\mu)} \frac{1}{g_{i_0 j}^2} & \sim & 3(N+M_{i_0 j}) - 2 N (1 - \gamma_{A} - 
\gamma_{B}) \\ \nonumber
\frac{d}{d\, \log (\Lambda/\mu)} \frac{1}{g^2} & \sim & 3N - 2 (N+M_{i_0 j}) (1 - \gamma_{A} - 
\gamma_{B}) ,  \mbox{for fixed} \quad i_{0} \quad 
\mbox{and} \quad j=1,\cdots,l \
\end{eqnarray}
For each diamond $(i_0 j)$ which belongs to the  $i_0$-th row, the fields $A$ and $B$ which enter 
in the previous equations correspond to the bi-fundamental representations
$(A_1)$ in $(\fund_{i_0,j},\antifund\; '_{i_0-1,j-1})$ of $SU(N+M_{i_0 j})_{i_0,j} \times SU(N)'_{i_{0}-1,j-1}$, 
$(A_2)$ in $(\fund_{i_0,j},\antifund\; '_{i_0,j})$ of  $SU(N+M_{i_0 j})_{i_0,j} \times SU(N)'_{i_0,j}$, 
$(B_1)$ in $(\fund\; '_{i_0-1,j},\antifund_{i_0,j})$ of $SU(N+M_{i_0 j})_{i_0,j} \times SU(N)'_{i_0-1,j}$ and
$(B_2)$ in  $(\fund\; '_{i_0,j-1},\antifund_{i_0,j})$ of  $SU(N+M_{i_0 j})_{i_0,j} \times SU(N)'_{i_0,j-1}$
where $SU(N)'_{i,j}$ represents the gauge group on the $(i,j)$ box and we use the fact that the 
 $(i_0,j), (i_0,j-1), (i_0-1,j)$ and $(i_0-1,j-1)$ boxes are adjacent to the $(i_0, j)$ diamond. 
The same for the $(i, j_0)$ diamond, we obtain the formulas:
\begin{eqnarray}
\label{rg12}
\frac{d}{d\, \log (\Lambda/\mu)} \frac{1}{g_{i j_0}^2} \sim 3(N+M_{i j_0}) - 2 N (1 - \gamma_{A} - 
\gamma_{B}) \\ \nonumber
\frac{d}{d\, \log (\Lambda/\mu)} \frac{1}{g^2} \sim 3N - 2 (N+M_{i j_0}) (1 - \gamma_{A} - 
\gamma_{B}) \
\end{eqnarray}
where $\gamma$ are the anomalous dimensions of the 
fields $A_1, A_2, B_1, B_2$ and near the fixed point $\gamma$  close to $-1/4$. 
By subtracting the second equation from the first 
in both (\ref{rg11}) and (\ref{rg12}), we obtain either
\begin{equation}
\frac{1}{g_{i_0 j}^2} - \frac{1}{g^2} 
\sim M_{i_0 j}[3 + 2 (1-\gamma_A - \gamma_B)] \log (\Lambda/\mu)
\end{equation}
or
\begin{equation}
\frac{1}{g_{i j_0}^2} - \frac{1}{g^2} \sim M_{i j_0}
[3 + 2 (1-\gamma_A - \gamma_B)] \log(\Lambda/\mu)
\end{equation}
We use the identification of the spacetime radial coordinate $r$ with the field theory scale and we see 
that the Type IIB supergravity solution has reproduced the field theoretic beta function,
this establishing the gravity dual
of the logarithmic RG flow in the ${\cal N} = 1$ supersymmetric 
$\prod_{i=1}^{k}\prod_{j=1}^{l} SU(N+M_{ij}) \times SU(N)^{'kl}$ gauge theory on $N$ regular and 
$M_{ij}, i = 1\cdots,k, j=1,\cdots,l$ fractional 
D3 branes. The agreement is between $\frac{1}{g_{ij}^2 N} - \frac{1}{g^2 N}$ 
at order $M/N$ in the large $N$ limit.

In \cite{kto} the authors have obtained an analytic form for the gravitational RG flow in the gauged
5-d Supergravity in the case of ${\AdS5 \times \T11}$. 
Their study concerned the back-reaction of the metric and 
5-form fields. Their results could be generalized to our case with 
the difference that the local geometry around each of the
$kl$ singular points should be used instead of the global geometry. 

In the case $l = 1$, the orbifolded ${\Z_k \times \Z_l}$ conifold becomes a generalized
${\Z_k}$ conifold. The horizon of the generalized conifold is singular and we need to partially
resolve in order to obtain a smooth Einstein manifold horizon with $k$ singular points. 
The procedure is just a particular case of our general recipe but the field theory 
on the worldvolume of D3 branes is not chiral
in this case. 

\section{Acknowledgments}
We would like to thank Keshav Dasgupta and Prabhakar Rao
for stimulating
discussions. The work of K. Oh is supported in part
by NSF grant PHY-9970664.

\end{document}